\documentclass[fleqn,usenatbib]{mnras}

\usepackage{newtxtext,newtxmath}

\usepackage[T1]{fontenc}
\usepackage[utf8]{inputenc}
\usepackage{ae,aecompl}

\usepackage{graphicx}	
\usepackage{amsmath}	
\usepackage{amssymb}	
\usepackage[squaren,Gray]{SIunits}
\usepackage{numprint}

\newcommand{\rlight}{r_{\rm L}}
\newcommand{\Rs}{R_{\rm s}}

\newcommand{\me}{m_{\rm e}}

\title[Neutron star magnetic field estimates]{The illusion of neutron star magnetic field estimates}

\author[J. P\'etri]{
J. P\'etri,\thanks{E-mail: jerome.petri@astro.unistra.fr}
\\
Universit\'e de Strasbourg, CNRS, Observatoire astronomique de Strasbourg, UMR 7550, F-67000 Strasbourg, France.
}

\date{Accepted XXX. Received YYY; in original form ZZZ}

\pubyear{2019}

\begin{document}
\label{firstpage}
\pagerange{\pageref{firstpage}--\pageref{lastpage}}
\maketitle

\begin{abstract}
Neutron stars radiate in a broad band spectrum from radio wavelengths up to very high energies. They have been sorted into several classes depending on their respective place in the $P-\dot{P}$ diagram and depending on spectral/temporal properties. Fundamental physical parameters such as their characteristic age and magnetic field strength are deduced from these primary observables. However this deduction relies mostly on interpretations based on simple vacuum or force-free rotating dipole models that are unrealistic. In this paper, we show that the computation of the stellar surface magnetic field is poorly estimated or even erroneous if multipolar components and particle loading are neglected. We show how quadrupolar magnetic field and monopolar winds alter field estimates and characteristic ages in the $P-\dot{P}$ diagram. Corrections brought by general relativity are also discussed. We derive some important parameters of pulsar physics such as the wind Lorentz factor~($\gamma$) times the pair multiplicity~($\kappa$) to be around $\gamma\,\kappa \approx \numprint{e8}-\numprint{e10}$. Therefore, the standard magnetodipole radiation losses formula must be used with caution to reckon neutron star surface magnetic fields and related secular evolution parameters. Depending on models we found that all field strengths, both for magnetars and for pulsars lie below the quantum critical value of $B_{\rm c} \approx \numprint{4.4e9}$~\SIunits{\tesla}.
\end{abstract}




\section{Introduction}

Neutron stars show up into different classes of compact objects like isolated neutron stars, accreting pulsars, magnetars, radio pulsars and millisecond pulsars \citep{harding_neutron_2013}. These many classes of neutron stars originate from their electromagnetic activity like pair creation/acceleration and radiation within the magnetosphere. Of particular importance is the magnetic field they harbour, usually close to or even higher than the critical field predicted by quantum electrodynamics to be about $B_{\rm c} \approx \numprint{4.4e9}~\SIunits{\tesla}$ when the quantum of energy associated to the electron cyclotron frequency equals its rest mass energy. The strength and topology of this magnetic field dictates neutron star behaviour and their outcome as a special subclass of neutron stars. The simplest approach assumes that the field lines are dipolar and the magnetic moment overlaps with the centre of the star. However, such straightforward approximations face more and more difficulties to explain recent high quality multi-wavelength light-curves and spectra. Thus off-centred dipoles emerged as a better way to explain pair creation efficiency \citep{harding_pulsar_2011} or gamma-ray pulsar light-curves \citep{barnard_effect_2016, petri_polarized_2017, kundu_pulsed_2017}.
Moreover, contributions from multipolar components are also usually neglected with respect to the dipolar part because the dipole is assumed to be dominant. While this is true at large distances~$r$ from the star because of the $r^{-(2\ell+1)}$ decrease of the field strength, where $\ell$ is the order of the multipole, and when the respective components are of the same amplitude, this no more holds for larger multipolar components anchored in the stellar crust. \cite{petri_multipolar_2015} investigated in depth the impact of multipoles on sky maps, spindown luminosities and braking indices. Extension to general relativity was also given by \cite{petri_multipolar_2017} who basically found the same conclusions.

For a pure dipole rotating in vacuum, the braking index is around $n=3$. However, several pulsars show braking indices well below this $n=3$ fiducial value. A simple explanation would be to add a monopolar spindown luminosity arising from a relativistic magnetized wind, shifting the braking index between one and three or by change in the moment of inertia \citep{hamil_braking_2015}. These models have been challenged by \cite{archibald_high_2016} who discovered the first pulsar with a braking index higher than~3. Several explanations have then been given like increasing the inclination angle as introduced by \cite{beskin_spin-down_1984}, the torque exerted by a plasma-filled magnetic dipole model \citep{eksi_inclination_2016}, the presence of a magnetic quadrupole \citep{petri_multipolar_2015} or even gravitational waves that are also of quadrupolar nature \citep{araujo_gravitational_2016}. A new model suggested by \cite{tong_possible_2017} includes braking indices less and larger than three, reconciling the whole set of observations. We stress that these uncertainties in the secular braking of the star goes back to the problem of neutron star magnetosphere models. The magnetic field strength, its topology and the particle loading within the magnetosphere are so far largely unconstrained.

Magnetic field strength etimates come indirectly from equating the magnetic dipole losses to the rotational slowdown, resulting in fields of the order $B=\numprint{e7}-\numprint{e10}$~\SIunits{\tesla} for normal pulsars and $B=\numprint{e4}-\numprint{e5}$~\SIunits{\tesla} for millisecond pulsars. Millisecond pulsars being much older, the magnetic field has time to decrease significantly. For anomalous X-ray pulsars and soft gamma repeaters, a similar calculation gives $B=\numprint{e9}-\numprint{e11}$~\SIunits{\tesla}. This field is necessary to explain the bursts releasing a colossal energy of~$\numprint{e37}-\numprint{e39}$~\SIunits{\joule}. For accretion-powered X-ray pulsars, electron cyclotron lines imply a field strength of $B=\numprint{e8}-\numprint{e9}$~\SIunits{\tesla} as reported by \cite{truemper_evidence_1978} for Her~X-1 and by \cite{wheaton_absorption_1979} for 4U 0115+63. Observations of 4U 1907+09 and Vela X-1 by \cite{makishima_cyclotron_1999} confirmed the cyclotron resonance line. Harmonics of the fundamental frequencies have also been measured by \cite{santangelo_bepposax_1999-1} and by \cite{heindl_discovery_1999} for 4U 0115+63. For some isolated stars, surface thermal emission, interpreted as proton cyclotron resonance, results in a more intense field strength of $B=\numprint{e9}-\numprint{e10}$~\SIunits{\tesla}. It has also been claimed that magnetars possess very strong magnetic fields above the critical value $B_{\rm c}$. Here again, this picture has been challenged by the fact that low magnetic field magnetars have been found \citep{rea_outburst_2013,rea_3xmm_2014} as well as high surface B-field pulsars. $B_{\rm c}$ seems therefore not to play a central role in the pulsar/magnetar dichotomy. However possible wrong guesses of the magnetic field must not be discarded. Indeed if higher order multipoles and strong deviations from the vacuum or force-free rotating dipole are expected within and around neutron stars, both classes could be reconciled without segregation according to the magnetic field. We will indeed demonstrate that $B_{\rm c}$ is not discriminating between pulsars and magnetars.

In this paper, we emphasize the strong sensitivity of magnetic field estimates on its topology, dipole versus multipole and particle loading, vacuum versus force-free and on general relativistic effects. In Sec.~\ref{sec:MultipolarRadiation}, we remind the state of the art simulations of pulsar magnetospheres in the two limits of vacuum and force-free environment, pointing out the braking efficiency related to the spindown luminosity in several plasma regimes. In Sec.~\ref{sec:FieldEstimate} we show detailed results about magnetic field strength estimates from the above assumptions. Derived quantities are the characteristic age exposed in Sec.~\ref{sec:AgeCaracteristique} and braking indices discussed in Sec.~\ref{sec:IndiceFreinage}. There we also derived some important pulsar magnetospheric parameters like pair multiplicity and particle Lorentz factors. Conclusions are drawn in Sec.~\ref{sec:Conclusions}.

\section{Multipolar electromagnetic radiation and spindown}
\label{sec:MultipolarRadiation}

In this section we compile analytical and numerical results from previous works about neutron star spindown luminosities for several magnetic field topologies in vacuum and when available, in force-free magnetospheres, for flat and curved spacetimes. To simplify the discussion and interpretation, we neglect corrections to these spindowns arising from terms of the order $R/\rlight$ (the first correcting term in vacuum is second order $O(R^2/\rlight^2)$). $R$ represents the neutron star radius and $\rlight$ the light cylinder radius~$\rlight = c/\Omega$, with $c$ the speed of light and $\Omega$ the neutron star rotation rate. This is certainly valid for normal radio pulsars but also a good estimate for millisecond pulsars for which $R/\rlight \lesssim 0.1$. The precision should remain better than~1\%.

We start with a reminder about neutron star electromagnetic braking in flat spacetime for vacuum multipoles and force-free dipoles (detailed force-free multipole solutions are not yet available). Indeed, so far, no simulations have been performed for force-free multipoles, but the formal dependence on $\Omega$ and $B$ remains the same apart from a constant numerical factor and the fact that the aligned mode $(\ell,m)=(2,0)$ mode then also radiates (we extrapolate results from the dipole case, knowing that there the aligned mode $(\ell,m)=(1,0)$ also radiates). We then extend the discussion to the general-relativistic counterpart, showing an increase in the magnetic field strength for an observer at rest on the stellar surface.

\subsection{Minkowskian case}

Plenty of results and literature exist about electromagnetic radiation in vacuum and matter when gravity is neglected. A brief overview of the main results useful to our present work is given below for vacuum radiation and force-free radiation for any multipole if available.

\subsubsection{Vacuum dipole}

The most studied rotator is a magnetic dipole radiating in vacuum. Exact solutions for a perfectly conducting sphere have been computed already by \cite{deutsch_electromagnetic_1955}, including corrections arising from the radial dependence according to spherical Hankel functions \citep{arfken_mathematical_2005}. There is no need to go into such details for a faithful estimate of magnetic field strengths. The formula for a point dipole is already sufficient. The well-known textbook result found in \cite{jackson_electrodynamique_2001} is
\begin{equation}
\label{eq:SpindownVacuumDip}
 L_{\rm dip} = \frac{8\,\upi}{3\,\mu_0\,c^3} \, B^2 \, \Omega^4 \, R^6 \, \sin^2 \chi
\end{equation}
where $\chi$ is the magnetic obliquity, $B$ the magnetic field strength at the magnetic equator, $\Omega$ the rotation speed and $R$ the neutron star radius. For a finite size magnet such as in \cite{deutsch_electromagnetic_1955}, there are corrections of the order $(R/\rlight)^2$ that we do not include. Note that an aligned rotator ($\chi=0\degree$) does not radiate and therefore the parallel component of the magnetic field~$B_\parallel$ is not constrained.  Note also that $B$ corresponds to the magnetic field value at the magnetic equator which is two times weaker than its value at the magnetic pole.

\subsubsection{Force-free monopole}

The force-free monopole represents the simplest model for a wind carrying energy to brake the star. Astonishingly, an exact solution for the full electromagnetic field has been found by \cite{michel_rotating_1973} and summarised in the spindown luminosity such as
\begin{equation}
\label{eq:SpindownMonopole}
 L_{\rm w}^{\rm mono} = \frac{8\,\upi}{3\,\mu_0\,c} \, B^2 \, \Omega^2 \, R^4.
\end{equation}
A more realistic field requires a split-monopole for which solutions for slow oblique rotators have been found by \cite{bogovalov_physics_1999}. The spindown luminosity is then the same as eq.~(\ref{eq:SpindownMonopole}). It does not depend on the obliquity~$\chi$. In the split monopole view, all particles emanating from the stellar surface contribute to the wind outflow. However, in a realistic geometry, the surface field is at least dipolar and only a small fraction of the surface area, actually the polar caps, contribute to the wind outflow. Thus eq.~(\ref{eq:SpindownMonopole}) cannot be used to computed the pulsar braking index. The  split-monopole solution only applies outside the light-cylinder. Inside the light-cylinder, the magnetic field is mostly dipolar and the realistic spindown is estimated through replacing $R$ by $\rlight$ and $B$ by $B(\rlight)$ which then shows the same formal dependence as for magneto-dipole losses in vacuum given by eq.~(\ref{eq:SpindownVacuumDip}). Therefore, we will not use this force-free monopole to compute losses due to a particle flow. We showed it for the completeness of our discussion.

Nevertheless, it is possible to estimated the spindown losses from such a wind by applying a trick similar to the one described above for finding the force-free dipole losses. Indeed, let us assume that, due to the wind, the force-free monopole or split monopole sets in at a radius $r_{\rm Y}$ less that $\rlight$. In eq.~(\ref{eq:SpindownMonopole}), replacing now $R$ by $r_{\rm Y}$ and $B$ by $B(r_{\rm Y})$ we get the force-free wind losses by
\begin{equation}
\label{eq:SpindownVent}
\dot E_{\rm w} = \frac{8\,\upi}{3\,\mu_0\,c}\, \frac{\Omega^2 \, B^2 \, R^6}{r_{\rm Y}^2} = {L_{\rm dip}^\perp} \, \left(\frac{\rlight}{r_{\rm Y}}\right)^2
\end{equation}
which varies between the force-free monopole losses for $r_{\rm Y}=R$ and the force-free dipole losses for $r_{\rm Y}=\rlight$.  The perpendicular spindown is define by $L_{\rm dip}^{\perp} = L_{\rm dip}(\chi=\upi/2)$. Eq.~(\ref{eq:SpindownVent}) is the same as eq.~(9) in \cite{harding_magnetar_1999} who derived it from a different perspective focusing on magnetars. Several very similar expression have been derived by other authors where contributions from wind and magnetic field are taken into account. See for instance \cite{tong_wind_2013}. Their reasoning is also explained in \cite{thompson_magnetohydrodynamics_1998}, we do not repeat it here.


The wind spindown in eq.~(\ref{eq:SpindownVent}) is always larger than the dipolar case whenever $r_{\rm Y}<\rlight$ (for a fixed magnetic field strength). In the opposite case if $r_{\rm Y} \gg \rlight$ the wind spindown becomes negligible and the braking is fully accounted for by the dipolar emission. This dependence is very similar to the results of a force-free aligned rotator computed by \cite{timokhin_force-free_2006} who also found a $r_{\rm Y}^{-2}$ dependence.

If field lines close already inside the light-cylinder, the usual size of polar caps must be replaced by $R_{\rm pc} \approx R\,\sqrt{R/r_{\rm Y}}$ (valid for $R\ll r_{\rm Y}$). Then assuming particles moving at almost the speed of light with Lorentz factor~$\gamma$ and taking a primary beam number density equal to the Goldreich-Julian density~$n_{\rm GJ} = 2\,\varepsilon_0\,\mathbf{\Omega} \cdot \mathbf{B}$, multiplied by the multiplicity factor~$\kappa$, the particle luminosity becomes
\begin{equation}
\label{eq:LuminositeVent}
 L_{\rm p} = 4 \, \pi \, \varepsilon_0 \, \Omega \, B \, c \, \frac{R^3}{r_{\rm Y}} \, \gamma \, \kappa \, \frac{\me \, c^2}{e} = \gamma \, \kappa \, \frac{c}{r_{\rm e}} \, e \, \Delta V
\end{equation}
where $\Delta V = \Omega \, B \, R^3/r_{\rm Y}$ is the potential drop between the centre and the rim of a polar cap and $r_{\rm e}$ the classical electron radius. In relativistic magnetized jets, longitudinal currents are expected to remain close to the Goldreich-Julian current density \citep{beskin_magnetohydrodynamic_2010}. The secondary plasma generated by pair cascade usually does not intervene in the electromagnetic torque exerted on the star, therefore disregarding the pair multiplicity factor. However this picture seems to fail for an oblique and especially for a perpendicular rotator. We therefore keep~$\kappa$ in all expressions.

Inspecting eq.~(\ref{eq:SpindownVent}), assuming a time-independent location of the Y-point $r_{\rm Y}$ for a given pulsar, the dependence on $\Omega$ is the same as for the split monopole, meaning $\dot E_{\rm w} \propto \Omega^2$ but with a different proportionality factor. It looks like radiation from a kind of monopolar magnetized outflow. In this approximation, adding the contributions respectively from the particle wind and from the dipolar Poynting flux, the total spin-down luminosity reduces to an expression
\begin{equation}
\dot E = a \, \Omega^2 + b \, \Omega^4
\end{equation}
where $a$ and $b$ are constants depending on magnetic field strength and particle loading. This expression is similar to the combined wind and dipole spin-down expected for magnetars as reported by \cite{harding_magnetar_1999} where some duty cycle is incorporated in the picture, switching from wind dominated to electromagnetic field dominated losses. Such formal dependences on $\Omega$ are used later to compute the time evolution of the neutron star period, its age and its field strength. Actually, generally speaking, when two or more mechanisms contribute to the spindown (for instance magnetic and gravitational wave braking, particle and magnetic braking among others), the losses are cast into a formal expression like
\begin{equation}
\label{eq:DoubleCoupleGeneral}
\dot E = a \, \Omega^\mu + b \, \Omega^\nu
\end{equation}
where $\mu$ and $\nu$ are two reals. For such a law, the braking index~$n$ remains in the interval $[\textrm{min}(\mu,\nu), \textrm{max}(\mu,\nu)]$. The exponents $\mu$ and $\nu$ are derived from the physics of stellar braking and need not to be integers. It is also conceivable to add other contributions with additional power laws like $\Omega^\sigma$.

In the force-free regime, the particle wind output remains negligible. It contributes to the current but not the luminosity. Expression (\ref{eq:SpindownVent}) is thus a good approximation to the torque exerted on the neutron star. The microphysics is hidden in the location of the Y-point depicted by the transition radius~$r_{\rm Y}$ which depends on $B$, $\Omega$ and $\gamma$. To simplify the subsequent study in this regime, we assume that this radius depends on the spin rate according to a power-law. Using a geometric weighted average, we set
\begin{equation}
r_{\rm Y} = \rlight^\beta \, R^{1-\beta}
\end{equation}
where $\beta\in[0,1]$ in order to ensure that $r_{\rm Y}$ always lies between the neutron radius~$R$ (monopole spindown) and the light-cylinder radius~$\rlight$ (dipole spindown). Such expression has already been introduced by \cite{sturrock_period-age_1971}. 
The braking index associated to this force-free wind losses is thus $n=1+2\,\beta$, explaining any braking index in the interval $n\in[1,3]$. Similar prescriptions have been introduced by \cite{contopoulos_revised_2006} for another study of pulsar spindown and justifying the above prescription by the efficiency of magnetic reconnection at the Y-point.

\subsubsection{Force-free dipole}

The first 3D simulations of the oblique rotator in force-free plasmas has been computed by \cite{spitkovsky_time-dependent_2006} followed by several other computations by \cite{kalapotharakos_toward_2012} and \cite{petri_pulsar_2012}, confirming the magnetic field configuration. A good approximate formula for the spindown luminosity is given by
\begin{equation}
\label{eq:SpindownDipole}
 L_{\rm ffe} = \frac{3}{2} \, L_{\rm dip}^\perp \, ( k_1 + k_2 \, \sin^2 \chi )
\end{equation}
where the constants $k_1$ and $k_2$ are obtained by fitting the results of the numerical simulation outputs. To a good precision, they are given by $k_1 \approx k_2 \approx 1$.
The spindown dependence on obliquity~$\chi$ is reminiscent of the vacuum case except that now the aligned rotator also radiates because of the electron/positron pair wind leading to an electric current braking the star by producing a toroidal magnetic field of the same order of magnitude as the poloidal field when crossing the light-cylinder. In this regime, the full magnetic field strength~$B$ at the surface is constrained by the magneto-dipole losses. There is no more freedom to choose $B_\parallel$ arbitrarily. Kinetic and MHD simulation results are also available but our coarse estimate remains precise enough for applications in Sec.~\ref{sec:FieldEstimate}.

\subsubsection{Vacuum multipolar radiation}

Computing multipolar radiation field solutions is more involved since mulitpoles are made of several configurations with $\ell\geq2$ and $m\leqslant \ell$. However it is already sufficient to consider the quadrupolar fields to understand the impact of multipolar radiation on magnetic field estimates. In Sec.~\ref{sec:FieldEstimate}, we show compelling examples containing a dipole and a quadrupole magnetic field.

As in the dipole radiation, the aligned quadrupole does not radiate. There is no mean to constrain this component by observation of the magnetic braking. However the $m=1$ and $m=2$ modes radiate such that
\begin{equation}
  L_{\rm quad}^{m=1} = \frac{L_{\rm quad}^{m=2}}{10} = \frac{128\,\upi}{135\,\mu_0\,c^5} \, B^2 \, \Omega^6 \, R^8 .
\end{equation}
The factor 10 difference between the $m=1$ mode and the $m=2$ mode arises from our normalisation of the multipolar components. Following \cite{petri_multipolar_2015}, we enforced a constant total magnetic energy outside the star whatever the geometry of the quadrupole. Any quadrupole is therefore produced by the rotation on a sphere by imposing two angles~$\chi_1$ and $\chi_2$. It is therefore similar to the dipole case where the total magnetic energy outside the star is independent of the obliquity~$\chi$.
So far, there are no force-free quadrupole magnetosphere solutions available in the literature. Nevertheless, we guess that the Poynting flux of a force-free aligned quadrupole would be of the same order as the vacuum quadrupole spindown. For a quadrupole, $B$ is related to the maximum value at the surface but it is not $B_{\rm max}$ because of the complicated dependence of the spherical components according to the coordinates $(\theta,\phi)$ on the stellar surface. However, the exact relation for an aligned quadrupole is
\begin{equation}
  B_{\rm max} = \sqrt{10} \, B .
\end{equation}
For the other modes $m=1,2$, the increase in the effective surface field strength compared to the value entering in the spin-down is of the same order of magnitude, around a factor~3. Recall that the true maximum surface dipole field is larger by a factor~2 than the field strength~$B$ used in the spin-down formula eq.~(\ref{eq:SpindownDipole}). Thus to compare the weight of quadrupole with respect to dipole, we assume that the value of $B$ in the spin-down formula is accurate enough. A precise definition of the field strength at the surface when dipole and quadrupole are present is even more involved and we keep the values of~$B$ as above from the respective spin-down expressions. We will show that good analytical field strength guesses can be found within a factor less than two.

\subsection{General-relativistic case}

When gravity comes into play, the magnetic field measured locally sensibly deviates from its extrapolation to a distant observer because of space-time curvature. Physical quantities, that is those actually measured by an observer must be properly defined and normalized according to some convention. In order to lay down the correct interpretation of measuring fields, we tackle the simple problem of a static magnetic dipole in Schwarzchild space-time. Frame dragging effects can be included \citep{petri_multipolar_2017} but they remain usually weak even for millisecond pulsars.

\begin{figure}
    \centering
    \includegraphics[width=0.9\columnwidth]{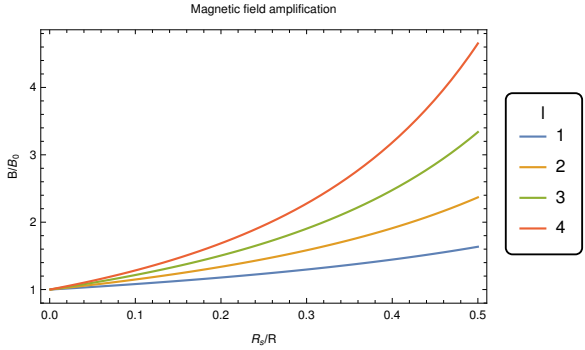}
    \caption{Surface magnetic field amplification~$B/B_0$ due to Schwarzchild spacetime compared to flat spacetime~$B_0$ for a mulipole of order~$\ell$ and depending on compacity~$\Rs/R$. The order~$\ell$ is shown in the legend.}
    \label{fig:AmplificationB}
\end{figure}

The 3+1 formalism already exposed in \cite{landau_classical_1971} but also in \cite{alcubierre_introduction_2008} enables us to write down Maxwell equations in a way similar to Euclidian space-time \citep{komissarov_3+1_2011}. In this approach, $\mathbf{D}$ and $\mathbf{B}$ are respectively the electric and magnetic field as measured locally by a fiducial observer. The exact solution for the general-relativistic static dipole in Schwarzchild space-time goes back to \cite{ginzburg_gravitational_1964}. It clearly shows an increase in the stellar surface field strength. Remind that for an aligned rotator the magnetic field is given by
\begin{subequations}
 \begin{align}
  B^{\hat r} & = - 6 \, \frac{B\,R^3}{\Rs^3} \, \left[ {\rm ln} \left( 1 - \frac{\Rs}{r} \right) + \frac{\Rs}{r} + \frac{\Rs^2}{2\,r^2} \right] \, \cos\vartheta \\
  \label{eq:MagneticStaticT}
  B^{\hat \vartheta} & = 3 \, \frac{B\,R^3}{\Rs^3} \, \left[ 2 \, \sqrt{ 1 - \frac{\Rs}{r}} \, {\rm ln} \left( 1 - \frac{\Rs}{r} \right) + \frac{\Rs}{r} \, \frac{2\,r-\Rs}{\sqrt{r\,(r - \Rs)}} \right] \, \sin\vartheta .
 \end{align}
\end{subequations}
$\Rs$ is the Schwarzschild radius of the star.
This expression is normalized such that a distant observer sees exactly a static dipole of strength~$B$ in vacuum and flat spacetime. The surface magnetic field amplification due to Schwarzchild spacetime compared to a flat spacetime dipole depending on compacity~$\Rs/R$ is shown in Fig.~\ref{fig:AmplificationB}. The increase goes up to 1.6 times for a compacity $\Rs/R=0.5$ as shown in solid blue line.

\subsubsection{Vacuum dipole}

Exact solutions to a slowly rotating magnetized neutron star are not easily found. Only some approximate solutions have been computed in the stationary regime by for instance \cite{rezzolla_electromagnetic_2004} and \cite{petri_general-relativistic_2013}. Time-dependent simulations were performed by \cite{petri_general-relativistic_2014} for a rotating dipole in general relativity. With respect to the energy losses by Poynting flux, frame dragging is negligible as shown by \cite{petri_multipolar_2017} for the dipole but also for multipoles. The spindown luminosity is the same as for the flat spacetime rotator. Corrections are only sensitive to unrealistically high rotation rates. Thus we use the expression~(\ref{eq:SpindownVacuumDip}) as a good guess. To get the correct value at the surface as measured by a fiducial observer, an amplification factor must be applied as shown in the plot in Fig.~\ref{fig:AmplificationB}.

\subsubsection{Force-free dipole}

Detailed simulations of general-relativistic force-free pulsar magnetospheres were done by \cite{petri_general-relativistic_2016} assuming a fixed dipole on the surface. Fitting formula for the spindown have also been recently established depending on compactnesses and rotation speed by \cite{carrasco_pulsar_2018}. \cite{ruiz_pulsar_2014} computed a full solution from the MHD stellar interior into the force-free external magnetosphere for several compactnesses. The spindown luminosities found are very similar. \cite{palenzuela_modelling_2013} included resistive effects into the MHD picture. To summarize the result, a good approximate expression is given by
\begin{equation}
 \label{eq:SpindownDipoleGR}
 L_{\rm ffe}^{\rm GR} = \frac{3}{2} \, L_{\rm dip}^\perp \, ( h_1 + h_2 \, \sin^2 \chi ).
\end{equation}
The coefficient $(h_1,h_2)$ depend on compactness and rotation rate. However for normal radio pulsars, the energy losses in eq.~(\ref{eq:SpindownDipoleGR}) tend to the flat spacetime expression given by eq.~(\ref{eq:SpindownDipole}). For millisecond pulsars, $h_1\approx1$ but $h_2\approx1.5$. The magnetic field strength is therefore not significantly affected by the discrepancy between normal and millisecond pulsars. We keep the flat spacetime formula as a good guess in general relativity too.

\subsubsection{Vacuum multipolar radiation}

The 3+1 formalism applied in \cite{petri_general-relativistic_2013} for a stationary dipole is used to compute numerical solutions of vacuum multipolar radiation fields in a slowly rotating spacetime. \cite{petri_multipolar_2017} gives results up to the $\ell=4$ octopole. Static multipole solutions in Schwarzchild spacetime are well known and expressed in terms of hypergeometric functions~${_2}F_1$ \citep{beskin_mhd_2009}. Fig.\ref{fig:AmplificationB} reports the magnetic field amplification for axisymmetric mulipoles with order $(\ell,0)$ up to the octupole~$\ell=4$. A monotonic increase in amplification is identified with respective maximum values of 1.6, 2.6, 3.4 and 4.6 for a compactness of 0.5. Curved spacetime amplifies higher~$\ell$ more than lower~$\ell$.

For the dipole, we saw that the flat spacetime approximation represents a good guess for slowly rotating general-relativistic dipoles. For quadrupoles and higher orders, the situation is more complex because the electric multipole of order~$\ell-1$ induced by the magnetic multipole of order~$\ell$ contributes more to the total Poynting flux whenever $m<\ell$. The $(\ell,m<\ell)$ multipoles radiate 2 to~3 times more than their flat spacetime counterparts for a compactness typically of 0.4-0.5. Thus the strength of a multipole of order~$(\ell,m<\ell)$ would be slightly less than the computation derived from Minkowskian multipoles.
For the sake of brevity, in the next section about magnetic field estimates, we only account for Minkowskian multipoles with possible magnetic amplification due to spacetime curvature. Nevertheless, this factor 2 to~3 for~$(\ell,m<\ell)$ must be kept in mind for slightly better guesses.

\section{Magnetic field estimates}
\label{sec:FieldEstimate}

In this section we show the impact of a relativistic magnetized outflow and a quadrupolar vacuum field on the line of constant stellar surface magnetic field~$B$ in the $P-\dot P$ diagram, emphasizing the discrepancy with respect to customary strength estimates. For the neutron star, we take a fiducial radius of $R=12$~\SIunits{\kilo\meter} and a moment of inertia of $I=\numprint{e38}~\SIunits{\kilo\gram \, \meter^2}$.

\subsection{Minkowskian case}

The magnetic field strength at the stellar surface is estimated assuming that the neutron star braking is entirely due to electromagnetic radiation through the Poynting flux. Within this approximation, the spindown is equated to the rotational kinetic energy~$E_{\rm kin}$ time derivative such that
\begin{equation}
 L_{\rm sd } = \frac{dE_{\rm kin}}{dt} = I \, \Omega \, \dot\Omega = - 4 \, \pi^2 \, I \, \dot P \, P^{-3} .
\end{equation}
Note that $L_{\rm sd }$ is negative implying indeed a braking of the star when $\dot P>0$.

\subsubsection{Pure vacuum or force-free dipole}

\begin{figure}
    \centering
    \includegraphics[width=0.95\columnwidth]{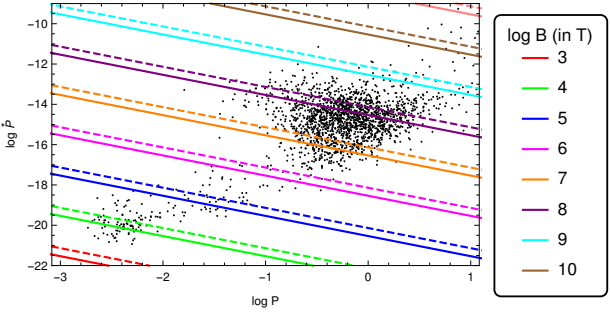}
    \caption{Lines of constant magnetic field found from the vacuum dipole losses (dashed lines) and force-free dipole losses (solid lines). The magnetic field strength is shown in the legend in a log scale.}
    \label{fig:DipolePPdot}
\end{figure}

For the vacuum dipole, the lines of constant perpendicular magnetic field~$B_\perp$ in the $P-\dot P$ diagram are given by
\begin{equation}
 \dot P = \frac{32\,\upi^3}{3\,\mu_0\,c^3} \, \frac{B^2_\perp \, R^6}{I\,P}
\end{equation}
where $B_\perp$ is the magnetic field strength perpendicular to the rotation axis. These
lines are shown in Fig.~\ref{fig:DipolePPdot}. It is the standard way to estimate magnetic field strengths in any class of neutron stars.

For a force-free dipole magnetosphere, we can constrain the full magnetic field because the aligned component also radiates. In such a case, the lines of constant magnetic field are obtained from an average spindown assuming an isotropic obliquity~$\chi$ distribution such that 
\begin{equation}
\frac{1}{4\upi} \, \int_0^\upi \int_0^{2\upi} (k_1 + k_2 \, \sin^2\chi)\,\sin\chi \, d\chi\,d\phi = k_1 + \frac{2}{3} \, k_2 \approx \frac{5}{3}.
\end{equation}
Thus the average force-free spin-down is
\begin{equation}
 L_{\rm ffe} = \frac{5}{2} \, L_{\rm dip}^\perp .
\end{equation}
The vacuum field is therefore $\sqrt{5/2} \approx 1.58$ times larger than the force-free field, on average, assuming the same spindown luminosity. The line of constant total magnetic field~$B$ field (recall that in plasma filled magnetospheres we constrain all components of the field) is now
\begin{equation}
 \dot P = \frac{80\,\upi^3}{3\,\mu_0\,c^3} \, \frac{B^2 \, R^6}{I\,P}
\end{equation}
and shown in Fig.~\ref{fig:DipolePPdot}. They are always slightly below the vacuum estimates but still of the same order of magnitude. There is no particular dependence on the period~$P$. All lines are straight and parallel to each other. This will no more be the case when a quadrupole component is added as shown now.

\subsubsection{Vacuum dipole + quadrupole}

\begin{figure}
    \centering
\includegraphics[width=0.95\columnwidth]{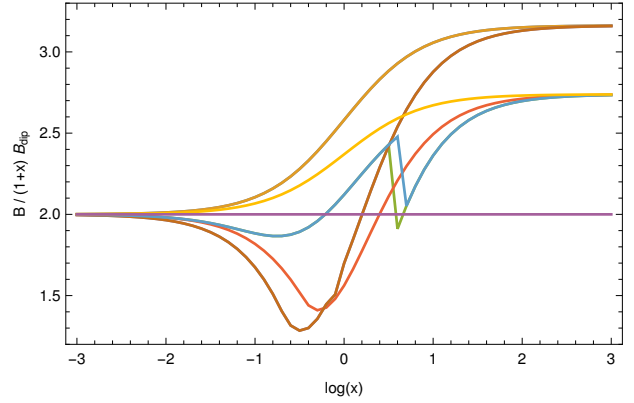}
\caption{Estimate of the maximum field strength at the surface compared to the guess given by $(1+x) \, B_{\rm dip}$ for different values of $x$ and the angles  $(\chi, \chi_1, \chi_2)$ set alternatively to $0$ and $\upi/2$. The constant line equal to 2 is shown for reference.}
\label{fig:Bmax}
\end{figure}

When a quadrupole is included, the situation gets more involved. In addition to weighting the respective contribution of the $m=1$ and $m=2$ quadrupolar components, we have to set their weights with respect to the dipolar part. Let us assume that the strength of the dipole field at the surface is $B_{\rm dip}$ and that of the quadrupole $B_{\rm quad}$. We introduce the ratio between both field strengths as 
\begin{equation}
 x = \frac{B_{\rm quad}}{B_{\rm dip}} .
\end{equation}
As a good guess, the total magnetic field at the surface is
\begin{equation}
\label{eq:Btot}
  B = B_{\rm dip} + B_{\rm quad} = (1+x) \, B_{\rm dip} .
\end{equation} 
The total magnetic field should take into account the $(\theta,\phi)$ dependence of each multipole but this is neglected in our discussion because it does not impact strongly on the estimate. Actually, the magnetic field strength depends on the spherical polar angles $(\theta,\phi)$. The above estimate is only a guess not the exact field strength at the surface. In order to get a better idea of the correctness of eq.~(\ref{eq:Btot}), we computed the exact analytical expression for the norm of any dipole/quadrupole configuration. According to \cite{petri_multipolar_2015}, the geometry is fully set by the ratio~$x$ and the three angles $(\chi, \chi_1, \chi_2)$. Detailed expressions can be found in the appendix of the same paper. By setting the angles alternatively to $0\degree$ and $90\degree$ and varying $x$, we looked for the maximum field strength at the surface. The result is then compared to eq.~(\ref{eq:Btot}) and shown in Fig.~\ref{fig:Bmax}. For a perfect agreement, we would expect the ratio $B/ (1+x) \, B_{\rm dip}$ to be equal to 2 (remembering that the polar field is twice the equatorial field for a dipole) for $x\ll1$ and close to~$\sqrt{10} \approx 3$ for $x\gg1$. We are always close to this value thus our guess is reasonable.

The spindown ratio then follows as
\begin{equation}
 \xi = \frac{L_{\rm quad}}{L_{\rm dip}} = \frac{16}{45}\,\eta \, x^2 \, \frac{R^2}{\rlight^2}
\end{equation}
where $\eta=1$ for the $m=1$ quadrupole and $\eta=10$ for the $m=2$ quadrupole. The dipolar field is then given by
\begin{equation}
 B_{\rm dip} = \sqrt{\frac{3\,\mu_0\,L_{\rm sd}\,\rlight^4}{8\,\upi\,c\,R^6\,(1+\xi)}} .
\end{equation}
The line of constant total magnetic field in the $P-\dot P$ diagram is defined by
\begin{equation}
  \dot P = \frac{32\,\upi^3}{3\,\mu_0\,c^3} \, \frac{B^2}{1+x} \, \frac{R^6}{I\,P} \, \left( 1 + \xi \right) .
\end{equation}

\begin{figure}
    \centering
    \includegraphics[width=0.95\columnwidth]{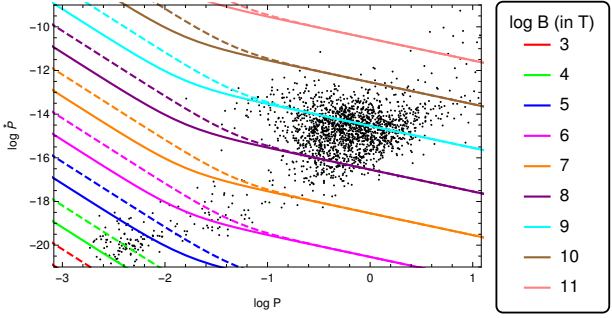}
    \caption{Lines of constant magnetic field from the vacuum dipole-quadrupole system with $x=100$ and for the mode $m=1$ in solid line and $m=2$ in dashed lines.}
    \label{fig:DipoleQuadrupolePPdot}
\end{figure}

For $x \ll 1$, the quadrupole does not contribute significantly to the total spindown luminosity. Indeed, the radiation field must be estimated from its value at the light cylinder. Knowing that a quadrupole decreases faster with distance than a dipole, it will become even weaker at the light-cylinder with $x_{\rm L} < x$. However, for $x\gtrsim1$, the quadrupole influences more the radiation field at the light-cylinder with respect to the dipole. If the light-cylinder is close to the stellar surface, thus for millisecond pulsars, it occurs already at moderate $x>1$. Moreover, if $x\gg 1$, the quadrupole component at the light cylinder cannot be neglected. Its emission produces sensitive effects at large distances as does the dipole emission. The transition from quadrupole luminosity dominance to dipole luminosity dominance arises around $L_{\rm quad} \approx L_{\rm dip}$ thus when the light-cylinder equals to
\begin{equation}
 \rlight = x \, R \, \sqrt{\frac{16}{45}\,\eta} .
\end{equation}
Translated into pulsar period, this gives respectively for $\eta=\{1,10\}$ the period expressed in milliseconds such that
\begin{subequations}
\begin{align}
P_{\rm m=1}(\textrm{in ms}) & = 0.150 \, x \\
P_{\rm m=2}(\textrm{in ms}) & = 0.474 \, x .
\end{align}
\end{subequations}
For $P < P_{\rm m}$ the quadrupole spindown of mode~$m$ is dominant whereas for $P > P_{\rm m}$ the dipole spindown is dominant. Consequently, for $x\lesssim10$ the quadrupole is never dominant whatever $P$ and $m$. For $x\gtrsim100$ the quadrupole starts to dominate in millisecond pulsar systems. This is clearly seen in the $P-\dot P$ diagram of Fig.~\ref{fig:DipoleQuadrupolePPdot} where the slope of constant $B$ changes around $P_{\rm m}$ from $-3$ to $-1$. Indeed, for a general multipole of order~$\ell$ the slope is given by $\dot P \propto B^2 \, P^{1-2\,\ell}$.
As a general rule, for high rotation rates, the higher multipole always dominates the spindown luminosity because of its higher power law dependence on $\Omega$. In the present case, the quadrupole dominates for millisecond pulsars whereas the dipole losses dominate for normal radio pulsars.

\subsubsection{Wind + FFE dipole}

The most common and realistic case is a combination of a FFE dipole and an ultra-relativistic particle wind carrying a significant fraction of the total spindown luminosity. The ratio between particle flux and Poynting flux is actually constrained by the braking index as we will show in Sec.~\ref{sec:IndiceFreinage}.

\begin{figure}
    \centering
    \includegraphics[width=0.95\columnwidth]{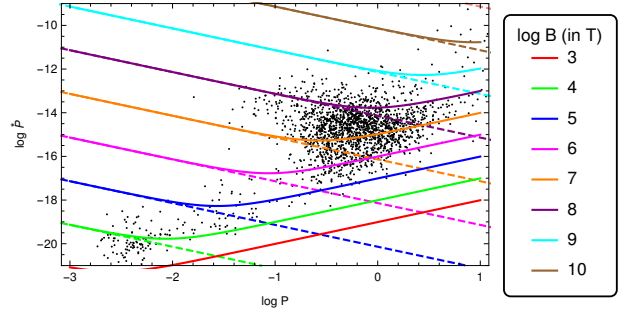}
    \caption{Lines of constant magnetic field from the wind force-free dipole system with $\gamma\,\kappa=10^8$.}
    \label{fig:VentDipoleFFEPPdot8}
\end{figure}

\begin{figure}
	\centering
	\includegraphics[width=0.9\columnwidth]{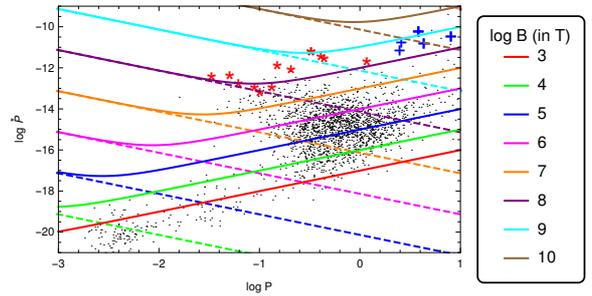}
	\caption{Corrected magnetic field strength for pulsars with measured braking index and parameters $\gamma\,\kappa=\numprint{e10}$. Pulsars with measured braking indices are depicted by red stars and magnetars with measured indices by blue crosses.}
	\label{fig:PulsarBCorrige}
\end{figure}

The line of constant magnetic field is found according to $\dot{E} = L_{\rm ffe} + L_{\rm p}$ for the simplest case $r_{\rm Y} = \rlight$. Thus
\begin{equation}
 \dot P = \frac{P^3}{4\,\upi^2\, I} \, ( L_{\rm ffe} + L_{\rm p}) .
\end{equation}
The FFE dipole Poynting flux is set by $B$ and $\Omega$. For the particle wind we need also a constrain on $\gamma\,\kappa$. This is guessed by the measured braking indices. Reasonable values would be $\gamma=10^{5-7}$ and $\kappa=10^{3-5}$. Thus we show two examples with parameters $\gamma\,\kappa=\{10^8,10^{10}\}$. Constant magnetic field lines in the $P-\dot P$ diagramm are shown for $\gamma\,\kappa=10^8$ in Fig.~\ref{fig:VentDipoleFFEPPdot8} and for $\gamma\,\kappa=10^{10}$ in Fig.~\ref{fig:PulsarBCorrige}. The presence of a luminous particle wind strongly alters the locii of constant magnetic field. Indeed, for normal radio pulsars, the presence of this wind significantly decreases the estimate of field strength by several orders of magnitude. This alteration is especially glaring for high $\gamma\,\kappa$ values as will be shown for pulsars and magnetars with measured braking indices in Fig.~\ref{fig:PulsarBCorrige}. 
In Sec.~\ref{sec:IndiceFreinage} we show how to get upper limits for $\gamma\,\kappa$ from braking index measurements.


\subsubsection{Force-free wind regime}

\begin{figure}
	\centering
	\includegraphics[width=0.95\columnwidth]{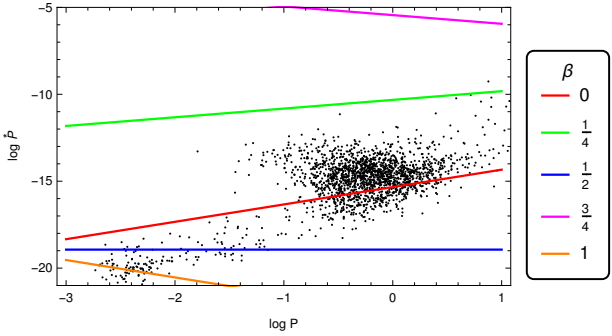}
	\caption{Lines of constant magnetic field from the force-free wind regime for $\beta\in\{0,1/4,1/2,3/4,1\}$ as reported in the legend. The field strength is $B=\numprint{e4}$~\SIunits{\tesla}.}
	\label{fig:FFEVentPPdot}
\end{figure}

\begin{figure}
	\centering
	\includegraphics[width=0.95\columnwidth]{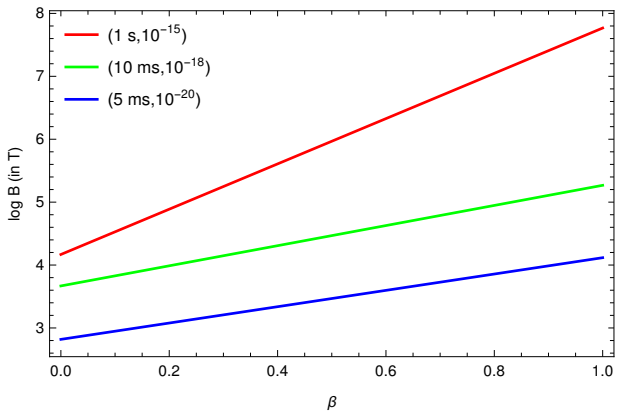}
	\caption{Magnetic field estimate variations due to the index ~$\beta$ for several $(P, \dot P)$ combinations.}
	\label{fig:BVentPPdot}
\end{figure}

Taking the view of a force-free wind regime where the Y-point is moved closer to the neutron star surface, we get another estimate of the magnetic field strength depending on the parameter~$\beta$. The lines of constant magnetic field are now defined in the $P-\dot P$ diagram by
\begin{equation}
\dot P = \frac{2 \, (2\,\upi)^{2+2\,\beta}}{3\,\upi} \, \frac{B^2 \, R^{4+2\,\beta}}{\mu_0 \, I \, c^{1+2\,\beta}} \, P^{1-2\,\beta} .
\end{equation}
Results are shown in Fig.~\ref{fig:FFEVentPPdot} for $\beta \in \{0,1/4,1/2,3/4,1\}$. It is clearly highlighted that the field estimate strongly depends on the value of $\beta$ thus on the location of the Y-point, and by many orders of magnitude. The cases $\beta=1/4$ and $\beta=3/4$ are extreme in the sense that they give field strengths far away from standard values. For a given pulsar with fixed $P$ and $\dot P$, the magnetic field variation is shown in Fig.~\ref{fig:BVentPPdot} for a young pulsar with $P=1$~\SIunits{\second}, $\dot P =\numprint{e-15}$ and for millisecond pulsars with $P=10$~\SIunits{\milli\second}, $\dot P =\numprint{e-18}$, $P=5$~\SIunits{\milli\second}, $\dot P =\numprint{e-20}$.

The magnetic field strength follows a law given by 
\begin{equation}
\log B = a +\frac{1}{2} \, \log \dot P + \beta \, \log (2\,\upi\,P\,R/c)
\end{equation}
where $a$ is a constant. The dependence on $\beta$ is only imprinted in the period~$P$ not in its derivative~$\dot P$ thus explaining the slopes observed in Fig~\ref{fig:BVentPPdot}.

\subsection{General-relativistic case}

What matters for the magnetic field strength is its value really measured by a local observer at the neutron star surface. Physics in gravitational fields then reduces to special relativity physics. The important point is to link this local measurements to signals received by a distant observer on Earth. This far away observer essentially measured the spin period and its derivative, $P$ and $\dot P$ respectively. Note that these quantities are measured in the asymptotic flat Minkowskian spacetime. The period felt by the local observer is smaller due to gravitational blue shifting. Thus when general relativity is included in the picture of a rotating magnetic dipole, care must be taken about space curvature and frame dragging effects. However \cite{petri_multipolar_2017} showed that frame dragging is negligible even for millisecond pulsars. Thus discrepancies in the estimates mainly arise from the curvature of space. 

As a general trend, the spindown luminosity measured by the distant observer in general relativity is enhanced compared to flat spacetime. Nevertheless, this effect vanishes for slowly rotating neutron stars for a dipole and for multipoles with $\ell=m$. For multipoles with $\ell>m$, the situation gets more involved due to efficient radiation of electric multipoles of lesser order equal to $\ell-1$. We neglect such complications introducing corrections to order unity. Consequently, the only relevant modification in the local magnetic field strength arises from the magnetic amplification explained in Sec.~\ref{sec:MultipolarRadiation}. For low order multipoles, this represents also only correcting factors of order unity, see Fig.~\ref{fig:AmplificationB}. As a rule of thumb, we keep estimates found in the Minkowskian case as good guesses to the actual field felt by the local observer.

\section{Characteristic ages}
\label{sec:AgeCaracteristique}

Related to the magnetic field strength and spindown luminosity is the characteristic age of significant magnetic braking. It is well known that for a single multipole of order~$\ell$ the braking scales as a power law of the spin rate~$\Omega$ such that $\dot \Omega = - k \, \Omega^n$ where $n=2\,\ell+1$ \citep{krolik_multipolar_1991, petri_multipolar_2015}. 
The characteristic age derived from this braking law, starting from a period at birth~$P_0$ going to the actual value~$P$ follows from
\begin{equation}
 \tau_{\rm c} = \frac{P}{(n-1)\,\dot P} \, \left[ 1 - \left(\frac{P_0}{P} \right)^{n-1} \right]
\end{equation}
for $n\neq1$. For short initial periods such that $P_0\ll P$, the characteristic age reduces to
\begin{equation}
\label{eq:AgeCaracteristiqueN}
\tau_{\rm c} = \frac{P}{(n-1)\,\dot P} .
\end{equation}
This expression is however not valid for a monopolar braking given by $n=1$. In the case of pure magneto-dipole loss ($n=3$), when the initial period is small, $P_0 \ll P$, the characteristic age becomes simply
\begin{equation}
 \tau_{\rm c }^{\rm dip} = \frac{P}{2\,\dot P} .
\end{equation}
This is the standard expression used to compute the age of isolated pulsars.
When several multipoles are present at the stellar surface, the braking index and characteristic age strongly depend on the ratio between the multipolar spindown contributions. In this section, we explore the consequences of a particle/force-free dipole system and a dipole/quadrupole system.

For a pure vacuum dipole, taking into account the evolution of the inclination angle due to the electromagnetic torque the braking index is
\begin{equation}
n^{\rm vac}=3+2\,\cot^2\chi.
\end{equation}
Other non vacuum models predict different evolutions of the inclination angle as summarised in \cite{beskin_radio_2018}. For instance in the Beskin-Gurevich-Istomin model (BGI) the braking index is approximately
\begin{equation}
n^{\rm BGI} \approx 1.93 + 1.5 \, \tan^2\chi
\end{equation}
whereas in the FFE/MHD regime it becomes
\begin{equation}
\label{eq:MHDIndice}
n^{\rm MHD} \approx 3 + 2 \, \frac{\sin^2 \chi \, \cos^2\chi}{(1+\sin^2\chi)^2} .
\end{equation}
We discuss in more detail the impact of the evolution of the inclination in onto the braking index in Sec.~\ref{sec:IndiceFreinage}.

\subsection{Wind and FFE dipole}

For a neutron star containing a monopolar wind and a FFE dipole, the braking law changes to
\begin{equation}
 \dot \Omega = - ( a_{\rm w} \, \Omega + b_{\rm ffe} \, \Omega^3 )
\end{equation}
where $a_{\rm w}$ and $b_{\rm ffe}$ are constants related to the wind and FFE dipole spindown luminosities respectively. For a spin rate of~$\Omega_0$ at birth, integration by separation of variables leads to the characteristic age by
\begin{equation}
 \tau_{\rm c}^{\rm w} = - \int_{\Omega_0}^{\Omega} \frac{d\Omega}{a_{\rm w}\,\Omega + b_{\rm ffe} \, \Omega^3} = \frac{1}{a_{\rm w}} \, \ln \left( \frac{\Omega_0}{\Omega} \, \sqrt{\frac{a_{\rm w} + b_{\rm ffe} \, \Omega^2}{a_{\rm w} + b_{\rm ffe} \, \Omega_0^2}} \right)
\end{equation}
from which follows the characteristic age as
\begin{equation}
 \tau_{\rm c }^{\rm w} = 2 \, \tau_{\rm c }^{\rm dip} \, (1+\xi) \, \ln \left( \eta \, \sqrt{\frac{1+\xi}{1+\xi\,\eta^2}} \right) .
\end{equation}
We introduced the ratio between FFE dipole and wind luminosities as
\begin{equation}
 \xi = \frac{b_{\rm ffe}\,\Omega^2}{a_{\rm w}} = \frac{L_{\rm ffe}}{L_{\rm w}} 
\end{equation}
and the ratio between initial and actual rotation rate as
\begin{equation}
\eta = \frac{\Omega_0}{\Omega} .
\end{equation} 
$\eta$ is usually assumed to be large, $\eta\gg1$.
The characteristic age diverges for a pure monopolar wind with $n=1$, being
\begin{equation}
 \tau_{\rm c}^{\rm w} = 2 \, \tau_{\rm c }^{\rm dip} \, \ln \eta
\end{equation}
whereas for a pure dipole $\xi \gg 1$ we retrieve the usual characteristic age of $\tau_{\rm c }^{\rm dip}$.

\begin{figure}
    \centering
    \includegraphics[width=0.95\columnwidth]{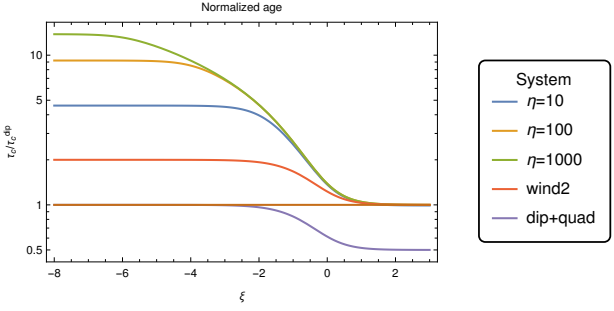}
    \caption{Characteristic age normalized to the pure dipole~$\tau_{\rm c}^{\rm dip}$ for any weight in the dipole-quadrupole system~$\xi$ and for a wind-dipole system with different spin ratios $\eta$.}
    \label{fig:AgeCaracteristique}
\end{figure}

The characteristic age normalized to the pure dipole~$\tau_{\rm c}^{\rm dip}$ and depending on $\xi$ and $\eta$ is shown in Fig.~\ref{fig:AgeCaracteristique} for $\eta=\{10,100,1000\}$. For reference, the pure dipole is depicted as the horizontal brown line. When the wind dominates the secular evolution of the neutron star, the characteristic ages can be underestimated by one order of magnitude. For the wind/FFE dipole system, the characteristic age is insensitive to $\eta$ as long as $\xi \gg 10^{-2}$. This is recognized by inspection of Fig.~\ref{fig:AgeCaracteristique}. This insensitivity will indeed applied to pulsars and magnetars studied in Sec.~\ref{sec:IndiceFreinage}.

The wind torque does not necessarily follow the $a_{\rm w}\,\Omega$ law. We could use any exponent as given in eq.~(\ref{eq:DoubleCoupleGeneral}). However, such expressions have no analytical form when integrated by separation of variable. Nevertheless, the general trend can be investigated by using another simple prescription for the wind+FFE spin-down such as
\begin{equation}
\dot \Omega = - ( a_{\rm w} \, \Omega^2 + b_{\rm ffe} \, \Omega^3 )
\end{equation}
where $a_{\rm w}$ and $b_{\rm ffe}$ are again constants related to the wind and FFE dipole spindown luminosities respectively. For a spin rate of~$\Omega_0$ at birth, integration by separation of variables leads to another characteristic age given by
\begin{equation}
\tau_{\rm c}^{\rm w} = - \int_{\Omega_0}^{\Omega} \frac{d\Omega}{a_{\rm w}\,\Omega^2 + b_{\rm ffe} \, \Omega^3} = \frac{1}{a_{\rm w}} \, \left( \frac{1}{\Omega} - \frac{1}{\Omega_0} \right) + \frac{b_{\rm ffe}}{a_{\rm w}^2} \, \ln \left[ \frac{\Omega \, (a_{\rm w} + b_{\rm ffe} \, \Omega_0)}{\Omega_0 \, (a_{\rm w} + b_{\rm ffe} \, \Omega)} \right]
\end{equation}
from which follows the characteristic age. For very high initial rotation rates $\Omega_0 \gg \Omega$ this age simplifies into
\begin{equation}
\label{eq:AgeCaracteristiqueVent} 
\tau_{\rm c }^{\rm w} = 2 \, \tau_{\rm c }^{\rm dip} \, (1+\xi) \, \left( 1 + \xi \, \ln \frac{\xi}{1+\xi} \right)
\end{equation}
with
\begin{equation}
 \xi =  \frac{b_{\rm ffe}\,\Omega}{a_{\rm w}} .
\end{equation}
The characteristic age varies between $P/2\,\dot P$ and $P/\dot P$ as expected from the standard age expression~(\ref{eq:AgeCaracteristiqueN}) by putting $n=3$ and $n=2$ respectively. This evolution is also depicted in Fig.~\ref{fig:AgeCaracteristique} and noted wind2.

\subsection{Dipole-quadrupole system}

For a neutron star containing only a vacuum dipole and a vacuum quadrupole, the braking law changes to
\begin{equation}
 \dot \Omega = - ( a_{\rm dip} \, \Omega^3 + b_{\rm quad} \, \Omega^5 ) .
\end{equation}
Integration by separation of variables leads to the characteristic age by
\begin{subequations}
\begin{align}
 \tau_{\rm c } & = -  \int_{\Omega_0}^{\Omega} \frac{d\Omega}{a_{\rm dip}\,\Omega^3 + b_{\rm quad} \, \Omega^5} \\
 & = \frac{1}{2\,a_{\rm dip}} \, \left( \frac{1}{\Omega^2} - \frac{1}{\Omega_0^2} \right) + \frac{b_{\rm quad}}{a_{\rm dip}^2} \, \ln \left( \frac{\Omega}{\Omega_0} \, \sqrt{\frac{a_{\rm dip} + b_{\rm quad} \, \Omega_0^2}{a_{\rm dip} + b_{\rm quad} \, \Omega^2}} \right) \nonumber .
\end{align}
\end{subequations}
For very high initial rotation rates $\Omega_0 \gg \Omega$ this age simplifies to
\begin{equation}
\label{eq:AgeCaracteristique}
 \tau_{\rm c } = \tau_{\rm c }^{\rm dip} \, (1+\xi) \, \left( 1 + \xi \, \ln \frac{\xi}{1+\xi} \right)
\end{equation}
by introducing now the ratio between quadrupolar and dipolar spindown as
\begin{equation}
 \xi = \frac{b_{\rm quad}\,\Omega^2}{a_{\rm dip}} = \frac{L_{\rm quad}}{L_{\rm dip}} .
\end{equation}
This is half the age given in eq.~(\ref{eq:AgeCaracteristiqueVent}) from the previous case.
For dominant dipolar spindown luminosity, $\xi \ll 1$, the characteristic age is approximately
\begin{equation}
 \tau_{\rm c } = \tau_{\rm c }^{\rm dip} \, ( 1 + \xi \, ( 1 + \log \xi )) .
\end{equation}
In the dominant quadrupolar spindown luminosity corresponding to $\xi \gg 1$, the characteristic age is approximately
\begin{equation}
\tau_{\rm c} = \frac{\tau_{\rm c }^{\rm dip}}{2} \, \left( 1 + \frac{1}{3\, \xi} \right)
\end{equation}
which indeed correspond to a $n=5$ braking index from a pure quadrupole.
The function in eq.~(\ref{eq:AgeCaracteristique}) is shown in violet in Fig.~\ref{fig:AgeCaracteristique} and noted dip+quad. It varies between $\tau_{\rm c }^{\rm dip}/2$ and $\tau_{\rm c }^{\rm dip}$. The error in the characteristic age determination is then at most a factor~2 when quadrupolar contributions are neglected.
Here also, the characteristic age varies between $P/4\,\dot P$ and $P/2\,\dot P$ as expected from expression~(\ref{eq:AgeCaracteristiqueN}) by putting $n=5$ and $n=3$ respectively.

Next, we show that knowledge about the braking index of pulsars and magnetars enables us to deeply constrain key magnetospheric parameters such as particle Lorentz factor and pair multiplicity, aside from the magnetic field although the results strongly depend on the underlying model.

\section{Effective braking index}
\label{sec:IndiceFreinage}

Measured braking indices significantly different from $n=3$ dipole could reveal the presence of non dipolar spindown. Indeed a dozen of pulsars have measured braking index gained from the knowledge of the spin second derivative according to
\begin{equation}
 n = \frac{\Omega \, \ddot \Omega}{\dot \Omega^2} .
\end{equation}
See \cite{archibald_high_2016} and \cite{hamil_braking_2015} for a recent compilation of pulsars and their braking index. All pulsars except one have $n\lesssim3$. Lower values are explained by the presence of a relativistic wind. Higher values are easily incorporated by adding a quadrupole. Therefore, starting with torques emanating from a monopole wind, a FFE dipole and quadrupole seems the easiest and most natural way to reconcile the whole set of data. So let us start with such a system.

\subsection{Wind+dipole+quadrupole system}

\begin{figure}
    \centering
    \includegraphics[width=0.9\columnwidth]{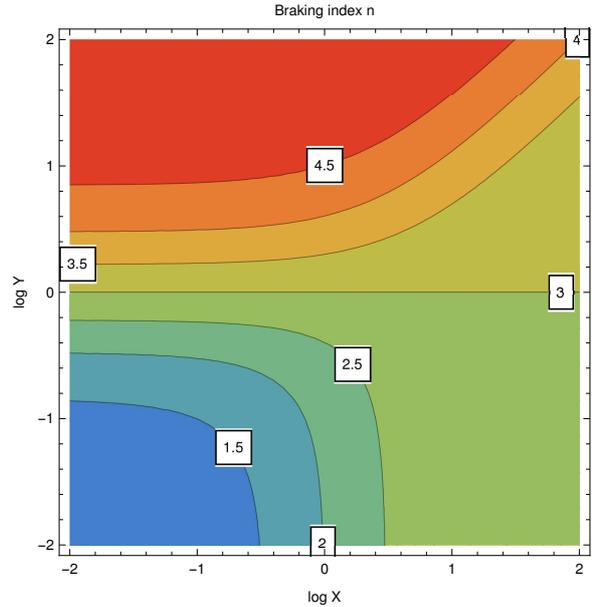}
    \caption{Isocontours of the braking index~$n$ for different ratio of dipole/wind and quadrupole/wind luminosities. The value of $n$ is shown in the black and white squares. This plot is for $p=1$.}
    \label{fig:IndiceFreinage}
\end{figure}

For a contribution from the wind, the dipole and the quadrupole, the braking law is advantageously written as
\begin{equation}
 \dot \Omega = - ( a \, \Omega^p + b \, \Omega^3 + c \, \Omega^5 )
\end{equation}
with $p<3$ the index of the wind torque depending on the microscopic dynamics of the particle acceleration in the polar caps. This spindown evolution leads to a braking index of
\begin{equation}
 n = \frac{p+3\,X+5\,Y}{1+X+Y}
\end{equation}
where we introduced respectively the ratio between dipole and wind luminosities and the ratio between quadrupole and wind luminosities as
\begin{subequations}
\begin{align}
 X & = \frac{b}{a}\,\Omega^{3-p} \\
 Y & = \frac{c}{a}\,\Omega^{5-p} .
\end{align}
\end{subequations}
The braking index is therefore always in the range $n\in[p,5]$.
\cite{alvarez_monopolar_2004} used the same law with explicit time dependent coefficient by performing a Taylor expansion of the general spindown law with $p=1$
\begin{equation}
 \dot \Omega = -F(\Omega,t) .
\end{equation}
The braking index is shown in Fig.~\ref{fig:IndiceFreinage}, depending on $\log X$ and $\log Y$. Interestingly, for $Y=\frac{3-p}{2}$ meaning that the quadrupole radiates almost as much as the monopolar wind, the braking index is always $n=3$ whatever $X$. Getting $n>3$ always requires a significant quadrupole torque at least as strong as the wind. 

For pulsars with known braking index less than~3, quadrupolar spindown is negligible. Thus for a pure wind+dipole system we set $Y=0$ and the ratio in spindown luminosities is directly related to the braking index through a relation similar to but generalizing the one found in \cite{lyne_45_2015}. We get
\begin{equation}
 X = \frac{n-p}{3-n} .
\end{equation}
Relevant parameters for pulsars with $n\in[1,3]$ are shown in Table~\ref{tab:PulsarParameters}. However, the measured braking indices are strongly affected by glitches as reported by \cite{espinoza_new_2017} who found short term evolution indices with $n>10$. 

Nevertheless, taking pulsars with measured averaged braking indices from Table~\ref{tab:PulsarParameters}, we found that the ratio lies in $0.1 \lesssim X \lesssim 10$ which means that spindown contribution from wind with $L_{\rm p} \propto \Omega^2$ and dipole are very similar. Assuming a particle outflow with Lorentz factor $\gamma$, the spindown eq.~(\ref{eq:LuminositeVent}) must be compared with the dipole. We use the force-free expression as the magnetosphere is now filled. This translates into a constrain on $\gamma\,\kappa$ via the ratio between wind and FFE dipole spindown as
\begin{equation}
\label{eq:RatioWindDipSpindown}
 \frac{L_{\rm w}}{L_{\rm ffe}} = \frac{3}{5} \, \gamma \, \kappa \, \frac{\me \, c^2}{e \, \Delta V} = \frac{1}{X}
\end{equation}
where $\Delta V = \Omega^2\,B\,R^3/c$ is the potential drop between the centre and the rim of a polar cap. In this particular model, we assume that $\rlight=r_{\rm Y}$ therefore $\beta=1$.
Note however that corrections from the wind are weak because of the location of these pulsars in the $P-\dot P$ diagram. They lie mostly in the transition zone between dipole and wind dominance, Fig.~\ref{fig:PulsarBCorrige}.
To compute the characteristic age, we set $\eta=100$ as it is insensitive to $\eta$ when $X \gg 10^{-2}$.

%

Several magnetars possess also measured braking indices as reported by \cite{gao_constraining_2016}. For those with $n\in[1,3]$, we can estimate $\gamma\,\kappa$ and the field strength in the same way as for pulsars. Even two GRB expected to harbor millisecond magnetars have measured braking indices from temporal evolution of their luminosity as reported by \cite{lasky_braking_2017}. Unfortunately, there is no period detection associated to these GRBs. Useful pertinent parameters are summarized in Table~\ref{tab:MagnetarParameters}.

\begin{figure}
	\centering
	\includegraphics[width=0.8\columnwidth]{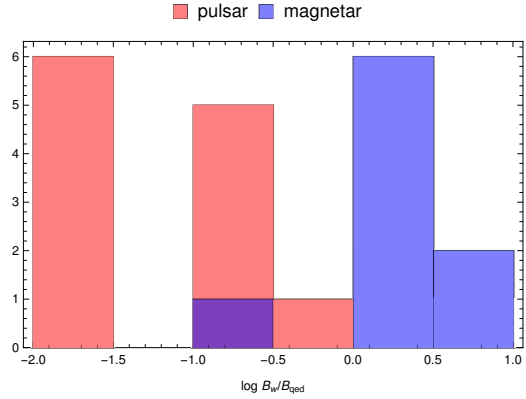}
	\caption{Magnetic field strength distribution~$B_{\rm w}$ for pulsars and magnetars corrected for the wind load. $B_{\rm w}$ is normalised to $B_{\rm qed}$.}
	\label{fig:HistBFFEreel}
\end{figure}

The histogram of magnetic field distribution according to the corrections brought by the wind are shown in Fig.~\ref{fig:HistBFFEreel} with the obvious bimodal distribution too, with magnetars possessing the highest fields $B\gtrsim\numprint{e9}$~\SIunits{\tesla} and the opposite for pulsars with  $B\lesssim\numprint{e9}$~\SIunits{\tesla}.

\begin{figure}
	\centering
	\includegraphics[width=0.8\columnwidth]{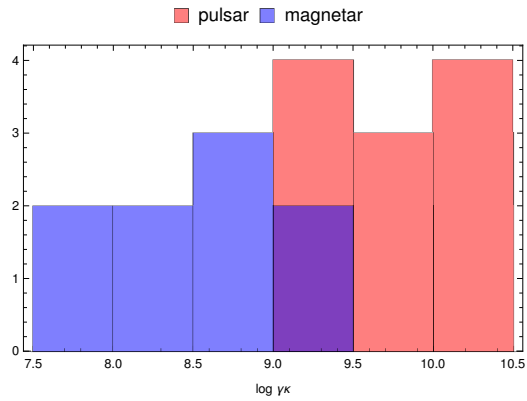}
	\caption{The product $\gamma\,\kappa$ distribution for pulsars and magnetars for the wind load.}
	\label{fig:HistGammaKappa}
\end{figure}

\begin{figure}
	\centering
	\includegraphics[width=0.8\columnwidth]{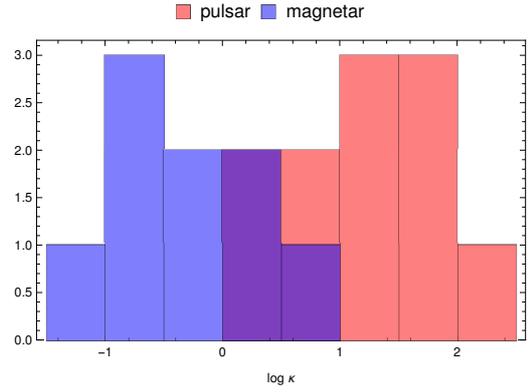}
	\caption{The pair multiplicity $\kappa$ distribution for pulsars and magnetars for the wind load.}
	\label{fig:HistKappa}
\end{figure}

\begin{figure}
	\centering
	\includegraphics[width=0.8\columnwidth]{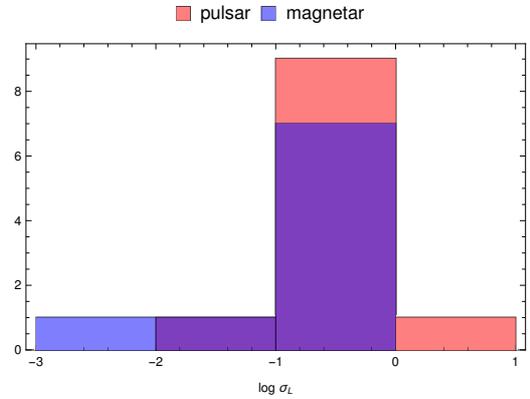}
	\caption{The magnetization parameter $\sigma_{\rm L}$ distribution for pulsars and magnetars for the wind load.}
	\label{fig:HistSigma}
\end{figure}

The Lorentz factor pair multiplicity product distribution $\gamma\,\kappa$ for pulsars and magnetars is shown in Fig.~\ref{fig:HistGammaKappa}. A bimodal distribution is readily seen with $\gamma\,\kappa$ for magnetars one to two orders of magnitude less than for radio pulsars. 
If the Lorentz factor is limited by the radiation reaction force due to curvature radiation taking curvature radius on the surface as $\rho_{\rm c} = \sqrt{R\,\rlight}$ and an electric field of the order $E=\Omega\,B\,R$ we get 
\begin{equation}
\gamma_{\rm CR} = \left( \frac{6\,\upi\,\varepsilon_0}{e} \, E \, \rho_{\rm c}^2 \right)^{1/4} .
\end{equation}
Therefore we can guess the pair multiplicity factor which is reported in the histogram of Fig.~\ref{fig:HistKappa}. The maximum Lorentz factor lies in the range $10^8-10^9$ irrespective of pulsar or magnetar.
The $\sigma$ parameter defined by the ratio of magnetic field energy density over particle energy density at the light-cylinder expressed as
\begin{equation}
\sigma_{\rm L} = \frac{B_{\rm L}^2}{\gamma\,\kappa\,n_{\rm L}\,\me \, c^2}
\end{equation}
is also computed and shown in Fig.~\ref{fig:HistSigma}. We did not found high $\sigma_{\rm L}$ values because the wind contributes significantly to the dynamics of the magnetosphere. We next switch to the force-free wind regime.

\subsection{Force-free wind regime}

\begin{figure}
	\centering
	\includegraphics[width=0.8\columnwidth]{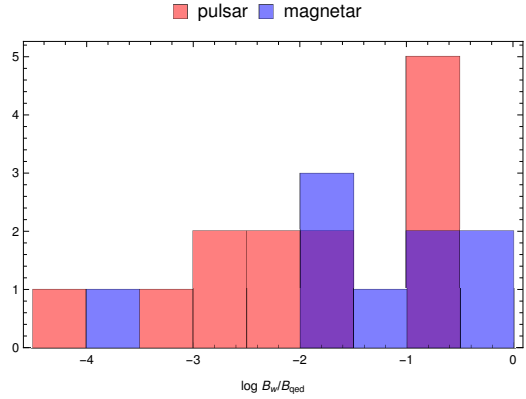}
	\caption{Magnetic field strength distribution~$B_{\rm w}$ for pulsars and magnetars for the force-free wind regime. $B_{\rm w}$ is normalised to $B_{\rm qed}$.}
	\label{fig:HistBreel}
\end{figure}

\begin{figure}
	\centering
	\includegraphics[width=0.8\columnwidth]{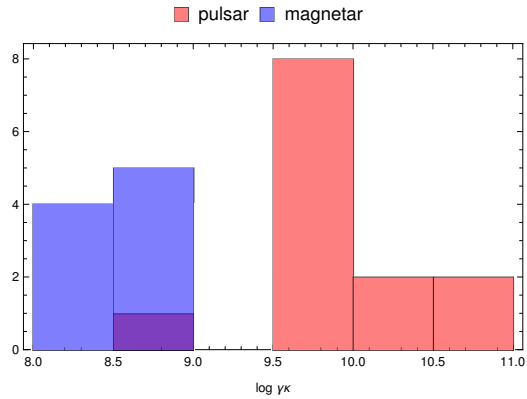}
	\caption{The product $\gamma\,\kappa$ distribution for pulsars and magnetars for the force-free wind regime.}
	\label{fig:HistGammaKappaFFEVent}
\end{figure}

\begin{figure}
	\centering
	\includegraphics[width=0.8\columnwidth]{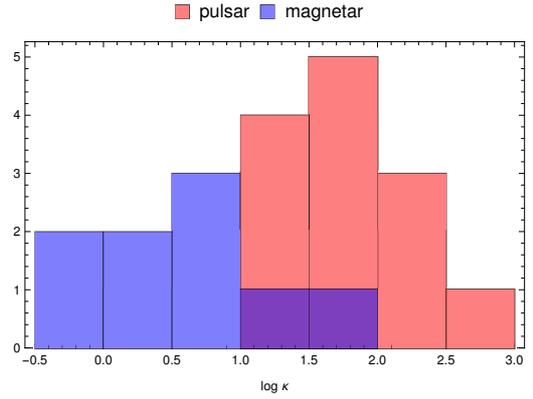}
	\caption{The pair multiplicity $\kappa$ distribution for pulsars and magnetars corrected for the force-free wind regime.}
	\label{fig:HistKappaFFEVent}
\end{figure}

\begin{figure}
	\centering
	\includegraphics[width=0.8\columnwidth]{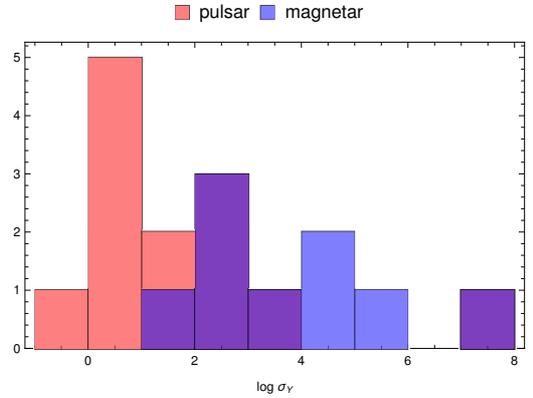}
	\caption{The magnetization parameter $\sigma$ distribution for pulsars and magnetars for the force-free wind regime.}
	\label{fig:HistSigmaFFEVent}
\end{figure}

Another prescription is given by the force-free split monopole beginning at a radius $r_{\rm Y}$, the location of the Y-point. We use eq.~(\ref{eq:SpindownVent}) to derive all important parameters of the pulsar magnetosphere. Its dependence on $\beta$ is directly related to the braking index by
\begin{equation}
\beta = \frac{n-1}{2}.
\end{equation}
Thus from observations, we immediately resolve for $\beta$, see table~\ref{tab:PulsarParameters}. Then the magnetic field strength is solution of $ \dot E = \dot E_{\rm w}$ thus
\begin{equation}
B = \sqrt{\frac{3\,\mu_0 \, c^{1+2\,\beta}}{8\,\upi} \, \frac{\dot E}{\Omega^{2+2\,\beta} \, R^{4+2\,\beta}}} .
\end{equation}
The magnetic field estimates in this model are given in the columns of Table~\ref{tab:PulsarParameters} noted by $B_{\rm w}$ and compared with the FFE dipole field~$B_{\rm ffe}$.
\begin{table*}
\centering
\begin{tabular}{lllllcccc}
	\hline
	PSR & $P$~(s) & $\dot P~(\numprint{e-12})$ & n & $\beta$ & $\log B_{\rm w}$ (T) & $\log B_{\rm ffe}$ (T) & $\log \gamma$ & $\log \kappa$ \\
	\hline
	J0534+2200 & 0.033392 & 0.4210  & 2.51$\pm$0.01   & 0.755 & 7.82 & 8.34 & 8.24 &  2.53 \\
	J0540-6919 & 0.050569 & 0.4789  & 2.140$\pm$0.009 & 0.57  & 7.46 & 8.45 & 8.03 &  2.49 \\
	J0835-4510 & 0.089328 & 0.1250  & 1.4$\pm$0.2     & 0.2   & 6.25 & 8.29 & 7.47 &  2.40 \\
	J1119-6127 & 0.40796  & 4.020   & 2.684$\pm$0.002 & 0.842 & 8.86 & 9.37 & 8.50 &  1.12 \\
	J1208-6238 & 0.440590 & 3.26951 & 2.598           & 0.799 & 8.69 & 9.34 & 8.42 &  1.11 \\
	J1513-5908 & 0.15125  & 1.531   & 2.839$\pm$0.001 & 0.919 & 8.72 & 8.94 & 8.54 &  1.52 \\
	J1640-4631 & 0.20644  & 0.976   & 3.15$\pm$0.03   & 1.075 & 9.13 & 8.91 & 8.75 &  1.01 \\
	J1734-3333 & 1.1693   & 2.279   & 0.9$\pm$0.2     & -0.05 & 5.62 & 9.48 & 6.86 &  1.96 \\
	J1803-2137 & 0.133667 & 0.13436 & 1.9             & 0.45  & 6.89 & 8.39 & 7.76 &  1.85 \\
	J1826-1334 & 0.101486 & 0.07525 & 2.2             & 0.6   & 7.16 & 8.20 & 7.94 &  1.71 \\
	J1833-1034 & 0.061883 & 0.2020  & 1.8569$\pm$0.001& 0.428 & 6.94 & 8.31 & 7.81 &  2.40 \\
	J1846-0258 & 0.32657  & 7.107   & 2.65$\pm$0.1    & 0.58  & 8.14 & 9.45 & 8.12 &  1.77 \\
	                            & & & 2.16$\pm$0.13   & 0.825 & 8.90 & 9.45 & 8.50 &  1.39 \\
	\hline
	\hline
\end{tabular}
\caption{Pulsar essential parameters with known braking index. Derived quantities are the luminosity ratio $L_{\rm ffe}/L_{\rm w}$, the pair multiplicity with maximal acceleration efficiency and the product $\gamma\,\kappa$ for less efficient acceleration mechanisms.}
\label{tab:PulsarParameters}
\end{table*}
The estimated magnetic field strengths are considerably reduces compared to the vacuum or force-free dipole estimates. We found that in this regime, it always lies below the quantum critical value as shown in the histogram of Fig.~\ref{fig:HistBreel} with magnetars ususally closer but still less than $B_{\rm c}$, see Table~\ref{tab:MagnetarParameters}.
\begin{table*}
\centering
\begin{tabular}{lllllcccc}
	\hline
	PSR & $P$~(s) & $\dot P~(\numprint{e-12})$ & n & $\beta$ & $\log B_{\rm w}$ (T) & $\log B_{\rm ffe}$ (T) & $\log \gamma$ & $\log \kappa$ \\
	\hline
	CXOU J1714& 3.8253 & 6.40 &    1.7  & 0.35  &  7.74 &  10.4 &  7.67 &  1.10 \\
							& & &  2.1  & 0.55  &  8.58 &  10.4 &  8.09 &  0.68 \\
							& & &  2.2  & 0.6   &  8.78 &  10.4 &  8.19 &  0.58 \\
	PSR J1622 & 4.3261 &  1.7 &    2.35 & 0.675 &  8.82 &  10.2 &  8.28 &  0.12 \\
							& & &  2.6  & 0.8   &  9.35 &  10.2 &  8.54 &  -0.14 \\
	SGR 0526  & 8.0544 &  3.8 &    1.82 & 0.41  &  7.85 &  10.5 &  7.71 &  0.45 \\	
						&  &  &    2.4  & 0.7   &  9.15 &  10.5 &  8.37 &  -0.19 \\
	SGR 1627  & 2.5945 &  1.9 &    1.87 & 0.435 &  7.84 &  10.1 &  7.81 &  0.95 \\
	Swift J1834 & 2.4823 & 0.796 & 1.08 & 0.04  &  6.08 &  9.91 &  6.97 &  1.62 \\
	\hline
	\hline
\end{tabular}
\caption{Magnetar essential parameters with known braking index. Derived quantities are the luminosity ratio $L_{\rm ffe}/L_{\rm w}$, the pair multiplicity with maximal acceleration efficiency and the product $\gamma\,\kappa$ for less efficient acceleration mechanisms.}
\label{tab:MagnetarParameters}
\end{table*}
To further estimate microscopic parameters, let us assume that this spindown is completely carried by the wind. This furnishes an absolute upper limit for $\gamma\,\kappa$ but does not reflect realistic cases as spindown is almost exclusively Poynting dominated. Nevertheless let us approximate the spindown such that $\dot E = L_{\rm p}$ from which we deduce the product $\gamma\,\kappa$ as
\begin{equation}
\gamma\,\kappa = \frac{2}{3} \,   \frac{e \, \Delta V}{\me \, c^2} = \sqrt{\frac{2}{3}} \, \sqrt{\frac{\dot E}{L_{\rm e}}}.
\end{equation}
Thus the product evolves simply as $\sqrt{\dot E}$. It is shown in Fig.~\ref{fig:HistGammaKappaFFEVent}. 
The pair multiplicity factor~$\kappa$ is also plotted in Fig.~\ref{fig:HistKappaFFEVent}.
The magnetization parameter $\sigma$ defined by the ratio of magnetic field energy density over particle energy density at the location of the Y-point
\begin{equation}
\sigma_{\rm L} = \frac{B_{\rm Y}^2}{\gamma\,\kappa\,n_{\rm Y}\,\me \, c^2}
\end{equation}
is shown in Fig.~\ref{fig:HistSigmaFFEVent}. Now the magnetization remains important $\sigma_{\rm Y} \gg 1$ for pulsars which is consistent with the force-free dynamics of the magnetosphere implied by the spindown luminosity eq.~(\ref{eq:SpindownVent}). However, for magnetars, this assertion is only marginally true. Certainly, in the force-free regime, the wind contribution to braking is low with $L_{\rm p}\ll \dot E$, therefore we expect $\gamma\,\kappa$ to be much lower than the values presented here.

It should be kept in mind that the numerical values presented in this study are only guesses and that more firm and precise prediction of the dynamics within the magnetosphere requires a better and deeper knowledge of matter radiation interaction which is not yet accessible neither analytically nor via computer simulations.

\subsection{Braking index and magnetic field profile}

Many mechanisms have been invoked to explain braking indices deviating from the fiducial $n=3$ value. All of them account for some detailed processes based either on moment of inertia changes, magnetic field decaying or increasing in time, obliquity variation or also precession. Based on some simple physical arguments, we remind that the spin-down luminosity is properly guessed by computing the Poynting flux emanating from of a sphere of radius~$\rlight$ in vacuum (or of radius $r_{\rm Y}$ in a force-free wind regime) with a proper choice of the magnetic field amplitude. Such arguments given by \cite{michel_theory_1991} and reused by \cite{petri_multipolar_2017} are applied to any poloidal magnetic field profile. Let us assume for instance that the field strength decreases like
\begin{equation}
 B_p = B \, \left( \frac{R}{r} \right)^s 
\end{equation}
where $s$ is some positive real number.
The field strength at the light-cylinder is therefore 
\begin{equation}
 B_L = B \, \left( \frac{R}{\rlight} \right)^s
\end{equation}
and the associated Poynting flux deduced again from the monopole solution starting at the light-cylinder is
\begin{equation}
 \mathcal{F} = \frac{8\,\upi}{3\,\mu_0}\,c \, B^2 \, R^2 \, \left(\frac{R}{\rlight}\right)^{2\,s-2} = \frac{8\,\upi}{3\,\mu_0} \, B^2 \, c^{3-2\,s} \, R^{2\,s} \, \Omega^{2\,s-2}
\end{equation}
leading to a braking index
\begin{equation}
 n = 2\,s - 3 .
\end{equation}
We retrieve the standard values for a monopole $(s=2, n=1)$, a dipole $(s=3, n=2)$ and generally for any multipole of order~$\ell$, $n=2\,\ell+1$. Consequently, any braking index can be explained by a proper radial decrease of the poloidal magnetic field. A non multipolar field decrease can be obtained by electric currents flowing within the magnetosphere. This requires non corotating magnetospheric models with electric currents flowing within the closed magnetosphere for instance.

A last important factor contributing to the braking index is the time evolution of the pulsar geometry, that is, in the dipolar case, the inclination angle evolution due to the electromagnetic torque. We discuss this topic in the following paragraph.

\subsection{Obliquity evolution and braking index}

So far we did not take into account the actual geometry of a particular pulsar. To simplify the discussion, we used a mean evolution by averaging against the obliquity~$\chi$ of the dipole. However, because the torque exerted on the neutron star depends on this angle~$\chi$, the braking index of a dipole in vacuum can significantly differ from the $n=3$ value when $\chi$ evolves on the same time scale as the period~$P$. Therefore, not only the stellar rotational braking evolution but also the obliquity evolution impacts on the braking index. Many prescriptions have been used to guess the time evolution of the two parameters~$\Omega$ and~$\chi$, depending on vacuum dipole or force-free rotator, the presence of a wind, itself depending or not on $\chi$ and lastly on possible acceleration gaps at work around the polar caps. We believe that all these models fall into a joint evolution of $\Omega$ and $\chi$ summarized by the following expressions
\begin{subequations}
\begin{align}
\dot \Omega & = - [(a+b\,\cos^2\chi)\,\Omega^\mu + (c+d\,\sin^2\chi)\,\Omega^\nu] \\
\Omega \, \dot \chi & = - (e\,\Omega^\mu + f \, \Omega^\nu) \, \sin\chi \, \cos \chi .
\end{align}
\end{subequations}
The dots indicate time derivatives. We introduced eight real parameters denoted by~$(a,b,c,d,e,f,\mu,\nu)$ (these parameters should not be confused with any physical constants like speed of light~$c$ and electric charge~$e$, used throughout the paper). Let us comment on these expressions. Two contributions are identified for the spindown and for the torque. The first term proportional to $\Omega^\mu$ arises from a wind carrying energy and angular momentum in the aligned case but decreasing in strength when moving to the perpendicular case, thus the presence of the $\cos^2\chi$ term. The second term proportional to $\Omega^\nu$ arises from a dipole, let it be vacuum or force-free, which is known to decrease the spindown removal and angular momentum when switching from a perpendicular to a aligned rotator, thus opposite to the wind contribution, explaining the $\sin^2\chi$ as seen in numerical simulations. The braking index then straightforwardly follows as
\begin{multline}
\label{eq:IndiceGeneral}
n = \frac{\mu\,(a+b\,\cos^2\chi)\,\Omega^\mu + \nu\,(c+d\,\sin^2\chi)\,\Omega^\nu}{(a+b\,\cos^2\chi)\,\Omega^\mu + (c+d\,\sin^2\chi)\,\Omega^\nu} \\
- 2 \, \frac{(b\,\Omega^\mu - d\,\Omega^\nu) \, (e\,\Omega^\mu + f \, \Omega^\nu) \, \sin^2\chi \, \cos^2 \chi}{[(a+b\,\cos^2\chi)\,\Omega^\mu + (c+d\,\sin^2\chi)\,\Omega^\nu]^2} .
\end{multline}
By appropriately choosing the eight parameters $(a,b,c,d,e,f,\mu,\nu)$, we retrieve many models discussed in the literature. However, we reduce the number of free parameters by noting that in the force-free dipole case, $c\approx d\approx f$ and $\nu=3$. For a vacuum magnetosphere we would also set $c=0$. For the wind contribution, we assume a similar formal dependence with $a\approx b \approx e$ and $\mu=1$, although other choices are possible. We add or remove a $\cos^2\chi$ dependence for the wind by putting $a=0$ or $b=0$ as done in several wind models. The first term on the right hand side of eq.~(\ref{eq:IndiceGeneral}) represents the braking index when the obliquity is assumed constant in time whereas the second term depicts the change in braking index induced by the evolution of the inclination angle~$\chi$. It can be positive or negative depending on the sign of $(b\,\Omega^\mu - d\,\Omega^\nu)$ associated to the dominant torque mechanism, wind or force-free.

As a simple representative case in this brief discussion, we choose $a=b=e$ and $c=d=f$, reminiscent of some vacuum or force-free model. Then, introducing the ratio between dipole and monopole spindown losses such as
\begin{equation}
\label{eq:SpindownRatio}
X = \frac{c}{a} \, \Omega^{\nu-\mu}
\end{equation}
the braking index reduces to a more tractable form given by
\begin{multline}
\label{eq:IndiceMuNu}
n = \frac{\mu\,(1+\cos^2\chi) + \nu\,(1+\sin^2\chi)\,X}{(1+\cos^2\chi) + (1+\sin^2\chi)\,X} \\
- 2 \, \frac{(1-X^2) \, \sin^2\chi \, \cos^2 \chi}{[(1+\cos^2\chi) + (1+\sin^2\chi)\,X]^2} .
\end{multline}
To go further, we assume a strictly monopolar wind and a strictly dipole force-free part thus setting $\mu=1$ and $\nu=3$. The time evolution of the rotation period and the angle, for a normalised rotation rate $\omega=\Omega/\Omega_0$ and a normalised time $\tau = 1/c\,\Omega_0^2$ is therefore 
\begin{subequations}
	\label{eq:Evolution}
	\begin{align}
	\dot \omega & = - \left[\frac{(1+\cos^2\chi)}{X_0} + (1+\sin^2\chi)\,\omega^2 \right] \\
	\dot \chi & = - \left[ \frac{1}{X_0} + \omega^2 \right] \, \sin\chi \, \cos \chi .
	\end{align}
\end{subequations}
$X_0$ is the initial ratio between dipolar and monopolar spindown as defined in eq.~(\ref{eq:SpindownRatio}).
When the dipole losses dominate, for $X\gg1$, we retrieve the force-free law given by eq.~(\ref{eq:MHDIndice}) which has a maximum at $n^{\rm FFE}_{\rm max} = 13/4$. In the opposite limit, for dominating monopole losses, for $X\ll1$, we get an index
\begin{equation}
\label{eq:MonopoleIndice}
n^{\rm mono} \approx 1 - 2 \, \frac{\sin^2 \chi \, \cos^2\chi}{(1+\cos^2\chi)^2} 
\end{equation}
which has a minimum index of $n^{\rm mono}_{\rm min}=3/4$.
\begin{figure}
	\centering
	\includegraphics[width=0.9\columnwidth]{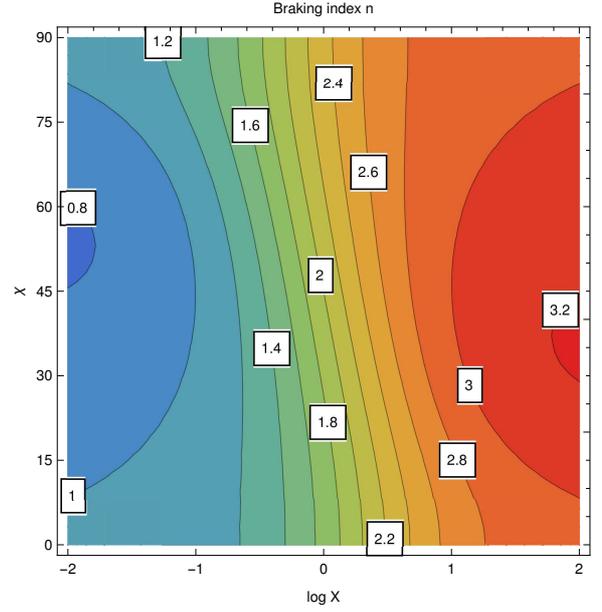}
	\caption{Braking index for a force-free+wind system with evolving inclination angle.}
	\label{fig:IndiceVent}
\end{figure}

Consequently, our simple prescription of a monopolar wind and dipolar FFE magnetosphere is able to reproduce all the measured braking indices from 0.9 to 3.15. Fig.~\ref{fig:IndiceVent} shows the range of braking indices depending on spindown ratio~$X$ and obliquity~$\chi$. For a dominant dipole, it lies around $n=3$, whereas for a dominant monopole it lies around $n=1$. In the region around equipartition $X\approx1$ the braking index evolves around $n=2$. These findings are insensitive to the angle~$\chi$.

\begin{figure}
	\centering
	\includegraphics[width=0.95\columnwidth]{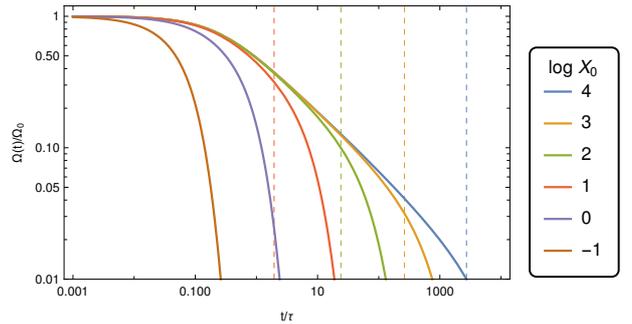}
	\caption{Evolution of the normalized pulsar angular velocity~$\Omega/\Omega_0$ for an initial obliquity~$\chi_0=60\degr$.}
	\label{fig:omega_evolution}
\end{figure}

\begin{figure}
	\centering
	\includegraphics[width=0.95\columnwidth]{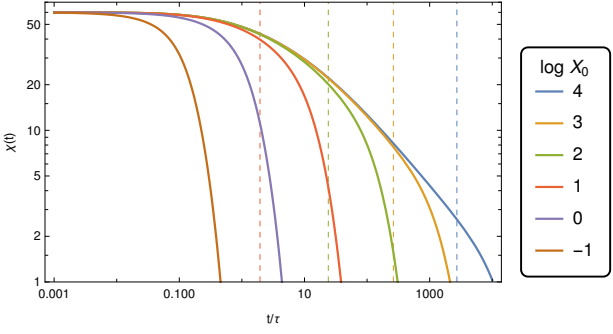}
	\caption{Evolution of the pulsar inclination angle~$\chi$ for an initial obliquity~$\chi_0=60\degr$.}
	\label{fig:chi_evolution}
\end{figure}

\begin{figure}
	\centering
	\includegraphics[width=0.95\columnwidth]{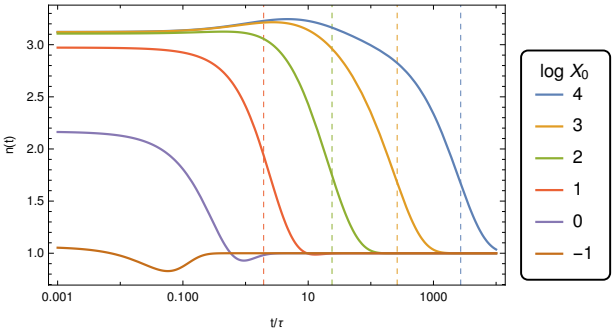}
	\caption{Evolution of the pulsar braking index~$n$ for an initial obliquity~$\chi_0=60\degr$.}
	\label{fig:indice_evolution}
\end{figure}

As a typical example of period and obliquity evolution, we numerically solved eq.~(\ref{eq:Evolution}) for a pulsar with initial obliquity~$\chi_0=60\degr$ and an initial rotation~$\Omega_0$. The evolution of the rotation rate~$\Omega(t)$ is shown in Fig.~\ref{fig:omega_evolution} for different initial ratio~$X_0$ from force-free domination~$X_0=10^4$ to wind domination~$X_0=10^{-1}$.
The related evolution of the obliquity~$\chi(t)$ is plotted in Fig.~\ref{fig:chi_evolution}.
In the regime of $X_0\gg1$ the evolution reduces to the MHD/force-free limit found in \cite{philippov_time_2014}. Nevertheless, when the star slows down, the initial ratio~$X_0$ decreases to lower values because $X(t)=X_0\, \omega(t)^2$ decreases too. Therefore, for very low angular velocities, the wind always dominates the dynamics when the transition from force-free to wind occurs, around $X_0\,\omega^2(t) \approx 1$. This condition is shown as a vertical dashed color line in the time evolution of the angular velocity Fig.~\ref{fig:omega_evolution}, obliquity Fig.~\ref{fig:chi_evolution} and braking index Fig.~\ref{fig:indice_evolution}. After that time, the velocity decreases faster and the shift towards alignment also accelerates. The inclination angle~$\chi(t)$ always evolves towards an alignment to $\chi = 0\degr$ when $\chi_0<90\degr$. Moreover correspondingly, the final braking index tends to the value $n=1$ because the spindown ratio $X(t)$ decreases towards the monopolar wind domination. Indeed, according to Fig.~\ref{fig:IndiceVent}, in the limit $X\ll1$ and $\chi\ll1$, the braking index is always $n=1$. This trend is clearly recognised in Fig.~\ref{fig:indice_evolution}.
This however does not mean that old pulsars will all tend to $n=1$ because when slowing down, the pair creation efficiency declines, implying an new increase in $X(t)$ and therefore and increase of the braking index towards $n=3$, according to Fig.~\ref{fig:IndiceVent} when evolving into the region $\chi\approx0\degr$ and $X\gg1$.

As a conclusion, we demonstrated the difficulty to explain the measured braking indices without including the temporal evolution of inclination angle~$\chi$. The fact that the product $\gamma\,\kappa$ is two orders of magnitude greater than its maximum possible value implies that the implicit assumption that the evolution of the inclination angle plays no role is probably far from reality. Moreover, there are several interpretations for a given value of the braking index. Consequently, we do not believe that fitting this braking index for any pulsar will be of any importance to support or not  pulsar slowing down models.

\section{Conclusions}
\label{sec:Conclusions}

Guessing confident values for the magnetic field of neutron stars is far from a trivial task. The sole knowledge about pulsar periods~$P$ and their corresponding derivatives~$\dot P$ is not enough to faithfully constrain the stellar surface field strength. Uncertainties come on one side from our ignorance of the particle load within the magnetosphere, that translates into number density, Lorentz factor and multiplicity, on the other side because of our ignorance of the magnetic topology at the surface. We demonstrated that constant magnetic field lines in the $P-\dot P$ diagram significantly differ from the standard vacuum or force-free dipole spindown approximation when multipoles and particle winds modify the flow and the energy balance. General relativity complicates even more this view by increasing the spindown efficiency and amplifying the local surface magnetic field strength. Under some simple assumptions, we showed how to compute the particle Lorentz factor and the pair multiplicity factor but giving too large contributions of the particle energy flux to explain braking indices $n<3$. This overestimate by two orders of magnitude takes its root in the assumption that the flux is particle dominated at the light cylinder and most importantly because we neglected the time evolution of the obliquity. Results about field strength and particle dynamics differ considerably depending on the underlying model. For instance a force-free wind regime decreases by several orders of magnitude the estimate of the true stellar magnetic field at the surface.

Our calculations rely on simple arguments that need to be refined with help of numerical MHD or kinetic simulations in order to better assess the interplay between particle dynamics and electromagnetic fields. It is only once this stage has been reached that we will be able to correctly guess realistic magnetic field measurements.

\section*{Acknowledgements}

I am grateful to the referee for helpful comments and suggestions. This work has been published under the framework of the IdEx Unistra and benefits from a funding from the state managed by the French National Research Agency as part of the investments for the future program. It also benefited from a CEFIPRA grant IFC/F5904-B/2018 and from the French National Research Agency (ANR) through the grant No. ANR-13-JS05-0003-01 (project EMPERE)







%
%
%
%
%


\bsp	
\label{lastpage}
\end{document}